\documentclass[12pt]{article}
\begin{document}

\begin{flushright}
UT-Komaba 00-3  \\
\end{flushright}

%%%%%%%%%%%%%%%%%%%%%%%%%%%%%%%%%%%%%%%%%%%%%%%%

\begin{center} 
{\Large{\bf  Pions in Lattice QCD with the Overlap Fermions at Strong
Gauge Coupling}}

\vskip 1.5cm

{\Large  Ikuo Ichinose\footnote{e-mail 
 address: ikuo@hep1.c.u-tokyo.ac.jp}and 
 Keiichi Nagao\footnote{e-mail
 address: nagao@hep1.c.u-tokyo.ac.jp}}  
\vskip 0.5cm
 
Institute of Physics, University of Tokyo, Komaba,
  Tokyo, 153-8902 Japan  
 
%\vfill
\end{center}

\vskip 3cm
\begin{center} 
\begin{bf}
Abstract
\end{bf}
\end{center}
In the previous paper we developed a strong-coupling expansion
for the lattice QCD with the overlap fermions and showed that
L\"usher's ``extended" chiral symmetry is spontaneously broken
in some parameter region of the overlap fermions.
In this paper, we obtain a low-energy effective action and 
show that there exist quasi-Nambu-Goldsone bosons which are
identified as the pions.
The pion field is a {\em nonlocal} composite field of quark 
and anti-quark even at 
the strong-coupling limit
because of the nonlocality of the overlap fermion formalism and
L\"usher's chiral symmetry.
The pions become massless in the limit of the vanishing bare-quark mass
as it is desired.

%%%%%%%%%%%%%%%%%%%%%%%%%%%%%%%%%%%%%%%%%%%%%%%%%%%%%%%%%%%%%%%%%%%%%%%%%
\newpage
One of the long standing problems in the lattice gauge theory
is the formulation of lattice fermions.
Recently a very promising formulation named overlap fermion
was proposed by Narayanan and Neuberger \cite{NN,Ne} and it has 
been studied intensively by both analytic and numerical methods.
In the previous paper \cite{IN} (which we shall refer to paper I hereafter),
we slightly extended the overlap 
fermion by introducing a ``hopping" parameter $t$ and studied
the lattice QCD by using both the $t$-expansion and the strong-coupling
expansion.
There we calculated the effective potential of the chiral condensation and
showed that L\"usher's extended chiral symmetry \cite{Lus}
is spontaneously broken at certain parameter region of the overlap fermion.
In this paper we shall obtain an effective action of low-energy excitations
and show that there exist quasi-Nambu-Goldsone bosons which are identified
as the pions.
As we show, the pion field is a {\em nonlocal} composite field 
of quark and anti-quark
even at the strong-coupling limit.

Action of the overlap fermion on the $d$-dimensional lattice
is given as follows,
\begin{equation}
S_F=a^d\sum_{n,m}\bar{\psi}(m)D(m,n)\psi(n),
\label{SF}
\end{equation}
where the covariant derivative $D(m,n)$ is defined as
\begin{eqnarray}
D&=&{1\over a}\Big(1+X{1 \over \sqrt{X^{\dagger}X}}\Big),  \nonumber  \\
X_{mn}&=&\gamma_{\mu}C_{\mu}(t;m,n)+B(t;m,n),  \nonumber   \\
C_{\mu}(t;m,n)&=&{t \over 2a}\Big[\delta_{m+\mu,n}U_{\mu}(m)-
\delta_{m,n+\mu}U^{\dagger}_{\mu}(n)\Big],  \nonumber  \\
B(t;m,n)&=&-{M_0\over a}+{r\over 2a}\sum_{\mu}\Big[2\delta_{n,m}
-t \delta_{m+\mu,n}U_{\mu}(m)-t \delta_{m,n+\mu}U^{\dagger}_{\mu}(n)\Big],
\label{covD}
\end{eqnarray} 
where $r$ and $M_0$ are 
dimensionless nonvanishing free parameters of the overlap lattice fermion 
formalism and $U_{\mu}(m)$ is gauge field on links.
Other notations are standard.
We have introduced a new parameter $t$.\footnote{As we explained in the
paper I, the $t$-dependence of the operator $D(m,n)$ is abosorbed
by a redefinition of $M_0$.}
The original overlap fermion corresponds to $t=1$.
For notational simplicity, we define 
\begin{equation}
A\equiv {1\over a}(dr-M_0), \;\; 
B\equiv {rt \over 2a},  \;\;
C\equiv {t\over 2a},
\label{ABC}
\end{equation}
and 
\begin{eqnarray}
\Gamma^-_\mu (m,n)&\equiv&\delta_{m+\mu,n}U_{\mu}(m)-
\delta_{m,n+\mu}U^{\dagger}_{\mu}(n),  \nonumber  \\
\Gamma^+_\mu(m,n)&\equiv&\delta_{m+\mu,n}U_{\mu}(m)+
\delta_{m,n+\mu}U^{\dagger}_{\mu}(n).
\label{-+}
\end{eqnarray}
In terms of the above quantities,
\begin{equation}
X_{mn}=A\delta_{mn}+C\sum\gamma_\mu \Gamma^-_\mu (m,n)-
B\sum\Gamma^+_\mu (m,n),
\label{X2}
\end{equation}
\begin{equation}
(X^\dagger)_{mn}=A\delta_{mn}-C\sum\gamma_\mu \Gamma^-_\mu (m,n)-
B\sum\Gamma^+_\mu (m,n).
\label{X3}
\end{equation}
From Eq.(\ref{ABC}), $B,C =O(t)$ and we consider $A=O(1)$ in the later
discussion.
Then it is rather straightforward to expand $D(m,n)$ in powers of
$t$,
\begin{eqnarray}
aD(m,n)&=&2\theta(A)\delta_{mn}
+{C \over |A|}\sum \gamma_\mu\Gamma^-_\mu(m,n)  \nonumber   \\
&& \;\;
   +{BC \over 2A|A|}\sum \gamma_\mu\Big(\Gamma^-_\mu(m,l)
   \Gamma^+_\nu(l,n)+\Gamma^+_\nu(m,l)\Gamma^-_\mu(l,n)\Big)  \nonumber   \\
&& \;\; +{C^2 \over 2A|A|}\sum\gamma_\mu\gamma_\nu \Gamma^-_\mu(m,l)  
   \Gamma^-_\nu(l,n)+O(t^3).  
\label{tD}
\end{eqnarray} 
Higher-order terms of $t$ are nonlocal and the $t$-expansion corresponds to 
a kind of the hopping expansion.
In the free field case or at the weak gauge coupling, the parameter region in
which fermion propagator has no species doublers is easily identified
in the $(M_0,r)$ plane.
However in the strong-coupling theory like QCD, the parameter region
of physical relevance should be determined by another requirement, because
the pole in the quark propagator is {\em not} a physical observable.
Therefore it is important to study the lattice QCD with the overlap
fermions in rather wide region of the parameter space. 

It is verified that the Ginsparg-Wilson (GW) relation \cite{GW}
\begin{equation}
D\gamma_5+\gamma_5D=aD\gamma_5D,
\label{GW}
\end{equation}
is satisfied by the $t$-expanded $D(m,n)$ in Eq.(\ref{tD}) at each order of
$t$.
Action of the fermion $S_F$ in Eq.(\ref{SF}) is invariant under 
the following extended chiral transformation discovered
by L\"uscher \cite{Lus},
\begin{equation}
\delta\psi(m)=\epsilon\gamma_5\Big(\delta_{nm}-aD(m,n)\Big)\psi(n),\;\;
\delta\bar{\psi}(m)=\epsilon\bar{\psi}(m)\gamma_5,
\label{extended}
\end{equation}
where $\epsilon$ is an infinitesimal transformation parameter.

Total action of the lattice QCD is given by
\begin{eqnarray}
S_{tot}&=&S_G+S_{F,M},  \nonumber   \\
S_G&=&-{1\over g^2}\sum_{pl}\mbox{Tr}(UUU^\dagger U^\dagger),  \nonumber  \\
S_{F,M}&=& S_F-M_B\sum\bar{\psi}(m)\psi(m),
\label{Stot}
\end{eqnarray}
where we have added the bare mass term of quarks.
We shall consider the strong-coupling limit in this paper though
a systematic strong-coupling expansion is possible.
We consider the $U(N)$ gauge group for large $N$.
It is easily verified that the following composite operators are
covariant under the transformation (\ref{extended})\cite{Nieder,Kiku},
\begin{equation}
\bar{q}(n)\equiv \bar{\psi}(n), \;\; q(n)\equiv \Big(1-{a\over 2}D(n,m)\Big)
\psi(n),
\label{qs}
\end{equation}
that is
\begin{equation}
\delta q(m)={\epsilon}\gamma_5 q(m), \;\; 
\delta \bar{q}(m)=\epsilon \bar{q}(m)\gamma_5.
\end{equation}
Hereafter we often set the lattice spacing $a=1$.
We consider the case of negative $A$ which is expected to
have desired properties of QCD \cite{IN}.

Partition function of the $U(N)$ QCD is given by the following functional
integral,
\begin{equation}
Z[J]=\int D\bar{\psi}D\psi DU \exp \Big\{-S_{tot}+\sum J(n)\hat{Q}(n)\Big\},
\label{partition}
\end{equation}
where $[DU]$ is the Haar measure and
\begin{eqnarray}
&& J(n)\hat{Q}(n)=J^\alpha_\beta(n)\hat{Q}_\alpha^\beta(n)  \nonumber  \\
&&  \hat{Q}_\alpha^\beta(n)=
{1\over N}\sum_aq_{a,\alpha}(n)
\bar{q}^{a,\beta}(n),
\label{Jm}
\end{eqnarray}
with color index $a$ and spinor-flavor indices $\alpha$ and $\beta$.
It should be remarked that the source $J$ is coupled to the nonlocal
operator $\hat{Q}$ instead of $\sum_a\psi_{a,\alpha}(n)
\bar{\psi}^{a,\beta}(n)$.
In Ref.\cite{IN-GN} we studied the gauged Gross-Neveu model
with the overlap fermions and showed that $\hat{Q}_\alpha^\beta(n)$
is the proper composite fields for the extended chiral symmetry, i.e.,
the order parameter is given by $\hat{Q}_\alpha^\beta(n)\delta_{\alpha\beta}$
and the Nambu-Goldstone bosons correspond to tr$_S(\hat{Q}(n)\gamma_5)$
where tr$_S$ is the trace over spinor indices.
In the rest of the present paper, we shall obtain the effective 
action $S_{eff}({\cal Q})$ defined as
\begin{equation}
Z[J]=\int D{\cal Q} e^{-S_{eff}({\cal Q})+J{\cal Q}}
\label{Seff}
\end{equation}
where integral over color-singlet {\em elementary} ``meson" field 
${\cal Q}^\alpha_\beta$
is defined as in Ref.\cite{IN}.

In order to evaluate the partition function (\ref{partition}),
we make a change of variables as
$$
(\bar{\psi},\psi) \Rightarrow  (\bar{q},q).
$$
Then the measure of the functional integral is transformed as
\begin{equation}
[d\bar{\psi}d\psi] \Rightarrow [d\bar{q}dq]e^{N\rm{Tr}\ln (1-{a\over2}D)},
\label{psi-q}
\end{equation}
where Tr in (\ref{psi-q}) is the trace over the spinor-flavor 
as well as the real-space indices. 
The contribution from the Jacobian $\mbox{Tr}\ln (1-{a\over2}D)$
is easily evaluated by the $t$-expansion.
In terms of $\bar{q}$ and $q$, 
\begin{eqnarray}
S'_{F}(q)&=&S_{F}(\psi) \nonumber \\
&=&\bar{q}(m)\Bigg[{C \over |A|}\sum \gamma_\mu\Gamma^-_\mu(m,n) 
+{BC \over 2A|A|}\sum \gamma_\mu\Big(\Gamma^-_\mu(m,l)
   \Gamma^+_\nu(l,n)+\Gamma^+_\nu(m,l)\Gamma^-_\mu(l,n)\Big)  \nonumber   \\
&& +O(t^3)\Bigg]q(n).
\label{Sq}
\end{eqnarray}
The term of $O((C/A)^2)$ in (\ref{tD}) has disappeared. As explained in 
paper I, this term is exactly determined from the term of $O(C/A)$
in (\ref{tD}) through the GW relation.
We expect that a similar phenomenon occurs for higher-order terms
of the $t$-expansion
which are determined by lower-order terms by the GW relation. 
That is, the terms of higher-order of $t$ which are determined 
through the GW relation 
drop when the action is rewritten in terms of $q$ and $\bar{q}$.
The mass term is also given by
\begin{eqnarray}
M_B\bar{\psi}(m)\psi(m)&=&M_B\bar{q}(m)\Big[\delta_{mn}-
{C \over 2A}\sum \gamma_\mu\Gamma^-_\mu(m,n) \nonumber  \\
&&-{BC \over 4A^2}\sum \gamma_\mu\Big(\Gamma^-_\mu(m,l)
   \Gamma^+_\nu(l,n)+\Gamma^+_\nu(m,l)\Gamma^-_\mu(l,n)\Big) \nonumber \\
&&   +O(t^3)\Big]q(n).
\end{eqnarray}
Total action of $q$ is given by
\begin{equation}
S'_{F,M}(q)=S_{F,M}(\psi).
\end{equation}

Integral over the gauge field can be performed by the one-link integral,
\begin{equation}
e^{W(\bar{D},D)}=\int dU_\mu \exp\Big[\mbox{Tr}(\bar{D}_\mu U_\mu
+U^\dagger_\mu D_\mu)\Big].
\label{onelink}
\end{equation}
Explicit form of the one-link integral $W(\bar{D},D)$ for the
present system is obtained as in paper I.
Gauge fields in $\Gamma^\pm_\mu(m,n)$ are replaced with composite
operators of $q$ and $\bar{q}$ after the integral over $U_\mu(n)$.
(For details see paper I.)
After some calculation, 
\begin{equation}
\int DUe^{-S'_{F,M}(q)+N\rm{Tr}\ln (1-{1\over2}D)+\sum J\hat{Q}}
=e^{-NM_B\mbox{tr}(\hat{Q})-S_2(\hat{Q})+\sum J\hat{Q}},
\label{U-int}
\end{equation}
where tr in (\ref{U-int}) is the trace over spinor-flavor indices and 
$S_2(\hat{Q})$ is some complicated function of $\hat{Q}$ which is obtained
in powers of $t$.
Let us define the following operators;
\begin{eqnarray}
\epsilon&=&\epsilon_\mu^{\delta\sigma}(m)=\Big({C \over A}\Big)^2
\Big(1-{M_B\over 2}\Big)^2\Big(\hat{Q}(m)\gamma_\mu\hat{Q}(m+\mu)
\gamma_\mu\Big)^{\delta\sigma},  \nonumber  \\
\epsilon' &=&\epsilon_\mu^{'\delta\sigma}(m)=\Big({C \over A}\Big)^2
\Big(1-{M_B\over 2}\Big)^2\Big(\hat{Q}(m+\mu)\gamma_\mu\hat{Q}(m)
\gamma_\mu\Big)^{\delta\sigma}.
\label{epsilons}
\end{eqnarray}
Then in terms of $\epsilon$'s, $S_2(\hat{Q})$ is given as\footnote{As 
we shall see, ${\cal Q} \sim O(t^{-1})$.} 
\begin{eqnarray}
{1\over N} S_2(\hat{Q})&=&-\sum_{m,\mu}\mbox{tr}
\Big[g(\epsilon_\mu(m))\Big]\nonumber \\
&&+{BC^3\over 2A^4}\Big(1-{M_B \over 2}\Big)^3 \nonumber  \\
&&\times \sum_{m,\mu,\nu}
\Bigg\{\mbox{tr}\Big[{\cal Q}(m+\mu)\gamma_\mu g'(\epsilon_\mu(m))
{\cal Q}(m)\gamma_\mu{\cal Q}(m+\mu+\nu)\gamma_\nu g'(\epsilon_\nu
(m+\mu))\Big]  \nonumber  \\
&&-\mbox{tr}\Big[{\cal Q}(m+\mu+\nu)\gamma_\mu g'(\epsilon_\mu(m+\nu))
{\cal Q}(m+\nu)\gamma_\mu{\cal Q}(m+\mu)\gamma_\nu g'(\epsilon'_\nu
(m+\mu))\Big]  \nonumber  \\
&&+\mbox{tr}\Big[{\cal Q}(m)\gamma_\mu g'(\epsilon'_\mu(m))
{\cal Q}(m+\mu)\gamma_\mu{\cal Q}(m+\nu)\gamma_\nu g'(\epsilon_\nu
(m))\Big]   \nonumber  \\
&&-\mbox{tr}\Big[{\cal Q}(m+\nu)\gamma_\mu g'(\epsilon'_\mu(m+\nu))
{\cal Q}(m+\mu+\nu)\gamma_\mu{\cal Q}(m)\gamma_\nu g'(\epsilon'_\nu
(m))\Big]  \nonumber  \\
&&+\mbox{tr}\Big[{\cal Q}(m+\nu)\gamma_\nu g'(\epsilon_\nu(m))
{\cal Q}(m)\gamma_\mu{\cal Q}(m+\mu+\nu)\gamma_\mu g'(\epsilon_\mu
(m+\nu))\Big]   \nonumber  \\
&&+\mbox{tr}\Big[{\cal Q}(m+\mu+\nu)\gamma_\nu g'(\epsilon_\nu(m+\mu))
{\cal Q}(m+\mu)\gamma_\mu{\cal Q}(m+\nu)\gamma_\mu g'(\epsilon'_\mu
(m+\nu))\Big]  \nonumber   \\
&&-\mbox{tr}\Big[{\cal Q}(m)\gamma_\nu g'(\epsilon'_\nu(m))
{\cal Q}(m+\nu)\gamma_\mu{\cal Q}(m+\mu)\gamma_\mu g'(\epsilon_\mu
(m))\Big]  \nonumber  \\
&&-\mbox{tr}\Big[{\cal Q}(m+\mu)\gamma_\nu g'(\epsilon'_\nu(m+\mu))
{\cal Q}(m+\mu+\nu)\gamma_\mu{\cal Q}(m)\gamma_\mu g'(\epsilon'_\mu
(m))\Big] \Bigg\} \nonumber  \\
&&+{N_{sf} C^4\over 4A^4}\Big(1-{M_B \over 2}\Big)^2
\sum_{m,\mu}\mbox{tr}\Big[{\cal Q}(m+\mu)\gamma_\mu g'(\epsilon_\mu(m))
{\cal Q}(m)\gamma_\mu g'(\epsilon'_\mu
(m))\Big]  \nonumber   \\
&& +O(t^3), 
\label{S2}
\end{eqnarray}
where $N_{sf}$ is the dimension of the spinor-flavor index and 
\begin{equation}
g(x)=1-(1-4x)^{1\over 2}+\ln \Big[{1\over 2}(1+(1-4x)^{1\over 2})\Big].
\label{g(x)}
\end{equation}

Elementary meson fields ${\cal Q}$ and their functional integral 
are introduced as in the previous case \cite{IN,KS},
\begin{eqnarray}
\int d\bar{q}dq \exp\Big({1\over N}J^\beta_\alpha q^\alpha_a
\bar{q}^a_\beta\Big) &=& \Big(\mbox{det}J\Big)^N  \nonumber   \\
&=&\oint d{\cal Q}
\Big(\mbox{det}{\cal Q} \Big)^{-N}\cdot e^{J\cdot{\cal Q}},
\label{int-meson}
\end{eqnarray}
where the integral over ${\cal Q}$ is defined by the contour integral,
i.e., ${\cal Q}$ is polar-decomposed as ${\cal Q}=RV$ with positive-definite
Hermitian matrix $R$ and unitary matrix $V$, and 
$\oint d{\cal Q}\equiv \int dV$ with the Haar measure of U($N_{sf}$)\cite{KS}.
From (\ref{int-meson}), there appear additional terms like
$(N\mbox{Tr}\log {\cal Q})$ in the effective action.
Therefore the effective action is given by
\begin{equation}
S_{eff}({\cal Q})=N\sum_n\Big[\mbox{tr} \ln {\cal Q}(n)
+M_B \mbox{tr} {\cal Q}(n)\Big]
+S_2({\cal Q}),
\label{Seff2}
\end{equation}

Effective potential of the chiral condensate is obtained from
$S_{eff}$ in Eq.(\ref{Seff2}) by setting 
$$
Q^{\alpha\beta}(n)=v\delta_{\alpha\beta}.
$$
In paper I we obtained
\begin{equation}
v={|A| \over 2C}\sqrt{{2d-1 \over d^2}} +O(t^0)+O(M_B).
\label{VEV2}
\end{equation}
From the above result, we can expect that there appear quasi-Nambu-Goldstone
bosons, i.e., pions.
Naive expectation is that pions correspond to the composite
operators like $\bar{\psi}\gamma_5\psi$ as in the continuum.
We examined the effective action and found that there is no
gapless excitation in the channel $\bar{\psi}\gamma_5\psi$.
However we found that there are massless modes in the channel 
$\mbox{tr}_F\Big(\hat{Q}\gamma_5\Big)=\mbox{tr}_F\Big(q\bar{q}\gamma_5\Big)$.
It is straightforward to obtain the effective action of the pions
by inserting the following expression of ${\cal Q}$
into $S_{eff}$ in Eq.(\ref{Seff2}),
\begin{equation}
{\cal Q}(m)=ve^{i\gamma_5\phi_5(m)}.
\label{phi5}
\end{equation}
For example from (\ref{epsilons}) and (\ref{phi5}),
\begin{equation}
\epsilon_\mu \propto e^{-i\gamma_5\nabla_\mu\phi_5}.
\end{equation}
We obtain
\begin{equation}
S_{eff}|_{{{\cal Q}(m)=ve^{i\gamma_5\phi_5(m)}}}=
2^{\frac{d}{2}}N \Big[ C_\pi
\sum_{m,\mu}\mbox{tr}_F\Big(\nabla_\mu\phi_5(m)\Big)^2
-\frac{M_B v}{2} \sum_m\mbox{tr}_F\Big(\phi_5(m)\Big)^2 \Big],
\label{S-pion}
\end{equation}
where $C_\pi$ is some positive constant.
Therefore the fields $\phi_5$ are quasi-Nambu-Goldstone pions as expected.

If we introduce elementary ``meson" fields ${\cal M}$ as in paper I,
i.e., 
$$
{\cal M}^\beta_\alpha(m) \sim {1\over N}\sum_a\psi_{a,\alpha}(m)
\bar{\psi}^{a,\beta}(m)
$$
and parameterize them as 
$$
{\cal M}(m)=ve^{i\gamma_5 \tilde{\phi}_5(m)},
$$
it is shown that there appears additional mass term of $\tilde{\phi}_5$
which is {\em finite} for $M_B \rightarrow 0$.
This fact must be important for numerical studies of QCD with
the overlap fermions.

In this paper we consider the leading-order of $1/N$.
We expect that in the next-leading order of $1/N$ a mass term
of the flavour singlet meson will appear from the Jacobian in 
(\ref{psi-q}).
This is a solution to the $U(1)$ problem.
The next-leading order terms in $1/N$ is under study and results
will be reported in a future publication\cite{WP}.

Finally let us comment on the domain wall fermions at strong coupling.
Very recently lattice $U(1)$ gauge model with the domain-wall fermions was
studied in the strong-coupling limit by the Hamiltonian 
formalism \cite{DW}.
There an effective Hamiltonian for low-lying color-singlet
degrees of freedom is obtained by treating the terms proportional
to the gauge fields $U_\mu(m)$ as perturbations.
This idea is very close to ours in paper I and the present paper
though we employ the Lagrangian formalism.
From the effective Hamiltonian obtained there, 
it is concluded that domain-wall
fermions at strong coupling suffer both the doubling problem and 
explicit breaking of chiral symmetry.
Furthermore it is claimed that the result also applied to the 
overlap fermions.
However we do not think that this is the case.
First, the effective Hamiltonian obtained there is {\em local}.
We expect that by integrating over heavy modes of the domain-wall
fermions there appear {\em nonlocal} terms in the effective Hamiltonian
or action of the light fermions.
Therefore it is suspected that the effect of heavy fermions is {\em not}
properly taken into account in the calculation in Ref.\cite{DW}.
Second, in order to prove the ``equivalence" between the domain-wall
and overlap fermions, bosonic Pauli-Villars fields must be introduced
in the domain-wall fermion formalism \cite{WD-OL,Kiku}.
As a result of the introduction of the Pauli-Villars fields,
the GW relation is satisfied by the overlap Dirac operator
in the action of the light fermions.
However in Ref.\cite{DW}, the bosonic Pauli-Villars fields are
{\em not} included at all.
From the above reasons, we do not think that the result in Ref.\cite{DW}
is applicable to the overlap fermions.
Actually, the effective action obtained in this paper by integrating over
the gauge fields is nonlocal even in terms of the chiral-covariant fields
$q(n)$ and $\bar{q}(n)$ and also there exists the extended chiral symmetry
for vanishing quark mass which is regarded as axial symmetry for
small hopping parameter $t$.

%%%%%%%%%%%%%%%%%%%%%%%%%%%%%%%%%%%%%%%%%%%%%%%%%%%%%%%%%%%%%%%%%%%%

\end{document}